# ADVANCED SERVICE DATA PROVISIONING IN RoF-BASED MOBILE BACKHAULS/FRONTHAULS

Mikhail E. Belkin[1], Leonid Zhukov[1] and Alexander S. Sigov[1]

[1]Russian Technological University MIREA, Moscow, Russia
belkin@mirea.ru, Zhukov_l@mirea.ru, sigov@mirea.ru

## ABSTRACT

*A new cost-efficient concept to realize a real-time monitoring of quality-of-service metrics and other service data in 5G and beyond access network using a separate return channel based on a vertical cavity surface emitting laser in the optical injection locked mode that simultaneously operates as an optical transmitter and as a resonant cavity enhanced photodetector, is proposed and discussed. The feasibility and efficiency of the proposed approach are confirmed by a proof-of-concept experiment when optically transceiving high-speed digital signal with multi-position quadrature amplitude modulation of a radio-frequency carrier.*

## 1. INTRODUCTION

In recent years, with the maturing of 5G NR systems, the design of access networks (ANs) has acquired some significant changes. In particular, the centralized radio access network (C-RAN) that passed from the 4G-LTE network, where a fronthaul interface connects various small cells deployed as remote radio heads (RRHs) to a centralized macro-cell deployed as a baseband unit (BBU) [1, 2], was standardized [3] into the next generation RAN (NG-RAN). Following it, some newer functional blocks, such as a central unit (CU), a distributed unit (DU), and a radio unit (RU) were introduced that is considered in detail in [4]. The key reason for this transformation was the use of the Common Public Radio Interface (CPRI) with time division multiplexing (TDM) between the BBU and RRH [4]. This approach led to data transfer rates up to hundreds of Gbps, which makes this interface economically unjustified for 5G and beyond.

Figure 1 depicts a typical architecture of NG-RAN, where CU, DU, and a set of RUs are duplex connected via fiber-optics communication lines (FOCLs), while RUs and mobile user terminals (UTs) intercommunicate wirelessly.

Generally, as optical networks evolve to fulfil highly flexible connectivity and dynamicity requirements, and supporting ultra-low latency services, it is increasingly important that a NG-RAN also provides reliable connectivity and improved network resource efficiency. Collection of different types of data from various sources is necessary for applying automation techniques to network management. However, the network must also support the capability to extract knowledge and form perception for performance monitoring, troubleshooting, and maintains network service continuity over a wide range of elements at various levels. Such scalability and flexibility are particularly important for the wide area network, in particular, for streaming telemetry [5]. Moreover, an efficient optical performance monitoring (OPM) design should consider different scenarios including large-scale disaster [6], when a prompt reaction is needed but limited bandwidth is available now.

Projecting the above problem that is common for telecom networks, for the purpose of this paper, we can conclude that in the newer generation, ANs providing the function of low-cost real-time monitoring and the quality-of-service (QoS) metrics are a matter of great importance from the point of view of their maintenance. In this case, the issue can be solved by additionally introducing a special node into the DU circuit (see Fig. 1), which is responsible for the accumulation and processing of monitoring results. However, a more promising solution from the point of view of reducing the total cost and latency, in our opinion, might be the introduction of an additional function from the existing indispensable element of its circuitry, return transmission of the optical signal to the CU, and its processing there. A promising technique for implementing this approach through the simultaneous use of an optically injection-locked vertical cavity surface emitting laser (OIL-VCSEL) as a laser source and a resonant cavity enhanced photodetector (RCE-PD) was proposed in [7] for a bi-directional optical communication and recently developed by us [8] referred to microwave photonics circuits.

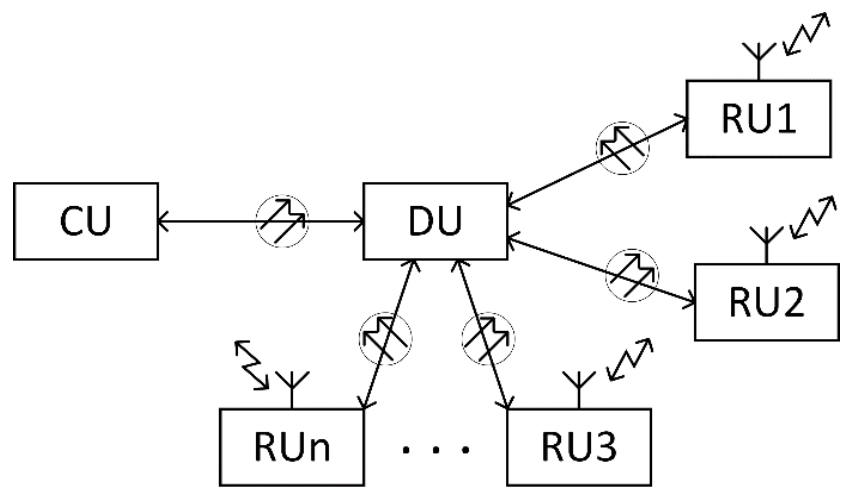

Figure 1. Conceptual block-diagram of 5G's NG-RAN

Elaborating the approach, in this paper after reviewing the modern service provisioning techniques in ongoing optical networks in Section 2, a newer design concept of OIL-VCSEL-based transmitter/receiver, which receives information data from CU via the downlink channel and simultaneously re-transmits them to RUs via the uplink channel for processing them at CU, is proposed and discussed in Section 3. The feasibility and efficiency of the proposed solution are confirmed in Section 4 by a proof-of-concept experiment, when optically transmitting a high-speed digital signal with 64-position quadrature amplitude modulation (QAM) of a 5-GHz radio frequency (RF) carrier. Section 5 concludes the paper.

## 2. MODERN TENDENCIES TO SERVICE PROVISIONING IN ONGOING NG-RANS

With the development of techniques and technologies of 5G NR networks, it became clear that this is not just a new standard for mobile communications. In general, the widespread worldwide introduction of core and access networks of the 5th and subsequent generations in the long term should transform our worldview and lead to a social transformation of the world community, radically changing the principles of communication, architecture, economy and the level of service of local and global telecom networks. For this purpose, known and newer for cellular communications concepts, paradigms, approaches, scenarios, technologies, mechanisms, tools and services are being developed. In particular, they include noted in Introduction NG-RANs,

Radio-over-Fiber (RoF)-based mobile backhauls/fronthauls, as well as Enhanced Mobile Broadband (eMMB) [9], Ultra-Reliable Low Latency Communication (URLLC) [10], Massive Machine-Type Communications (mMTC) [11], Internet of Everything [12], Slice-based Networks for Heterogeneous Environments [13], Software Defined Networking (SDN) [14], Network Function Virtualization (NFV) [15] and so on.

The results of the above innovations should lead to significant improvements in the QoS and key parameters of NG RANs. Thus, the recommendation 3GPP TR 38.913 identified the following outstanding key indicators of new generation networks:

- downlink peak data rate up to 20 Gbps with spectral efficiency 30 bps/Hz
- uplink peak data rate up to 10 Gbps with spectral efficiency 15 bps/Hz
- the minimum delay in the radio access subsystem for URLLC services is 0.5 ms, for eMBB services - 4 ms
- the maximum density of the IoT devices connected to the network in urban territories is 1'000'000 devices/sq. km
- autonomous operation of the IoT devices without recharging the battery for 10 years;
- vehicle mobility at a maximum speed of 500 km/h.

Along with these, a critical problem arose related to ensuring the above parameters during the maintenance of realistic NG-RANs through the development of advanced operational management's principles, approaches and schemes. In general, advanced monitoring framework of optical networks aimed at the continuous, remote, automatic and cost-effective supervision of the physical layer has to satisfy the basic set of the requirements:

(i) fast and accurate detection of the performance degradation and service disruptions
(ii) accurate tracking down location of the network failure
(iii) monitoring should be non-intrusive and not affecting normal network operation
(iv) compatible with various types of optical networks.

Besides, to meet the goals of 5G NG and beyond, network infrastructures should facilitate a high level of flexibility and automation. In particular, monitoring and data analytics give rise to estimate accurately the QoS of new light paths, to anticipate capacity exhaustion and degradations, or to predict and localize failures, among others to facilitate this automation [16]. At the same time, network operations and management (OAM) increasingly relies on the ability to stream and process in real-time data from network equipment. An integral part of the OAM is to make sure whether the operational conditions are normal or not and intervene, if needed, by quickly recovering and mitigating the occurred problems [17]. The goal to have network management automation and abstraction of open line systems (OLS) could be possible by the software defined network (SDN) technologies, which requires accurate OPM data from the elements of the network [18].

## 3. DESIGN CONCEPT

Based on the results of the review in Section 2, it can be unambiguously concluded that any existing approach to monitoring the QoS of a FOCLs leads to the complication of the DU circuitry and operation, and consequently to the increase in its cost and the latency of signal transmission via the AN. Therefore, the solution related to the introduction of an additional function to the existing indispensable element of its circuitry, namely the simultaneous use of an OIL-VCSEL as a laser source and a RCE-PD, is promising in principle.

Revealing the proposed concept, Figure 2 demonstrates a block diagram of a communication channel, containing the all three functional units of NG-RAN (see Figure 1). It is worth noting that the block-diagram presented in this Figure has three key distinguishing features in according to NG-RAN concept. First, at the CU, in order to simplify the circuitry of subsequent units, the conversion of the modulation format of the optical carrier from a baseband (BB) to QAM of RF subcarrier is performed so that the digital signal is then transmitted over FOCLs using a RF equal to the allocated frequencies of the corresponding RU [19]. Secondly, a duplex optical channel is introduced between the CU and DU, where the information signal is transmitted in the downlink direction, and the QoS data signal - in the uplink direction. Thirdly, in the DU, an OIL-VCSEL is connected through an optical circulator, where reflected optical output is transmitted via downlink to the RU, and the detected RF signal with added QoS data is again converted into the optical range using a standard low-cost optical transmitter and returned to the CU for the subsequent processing.

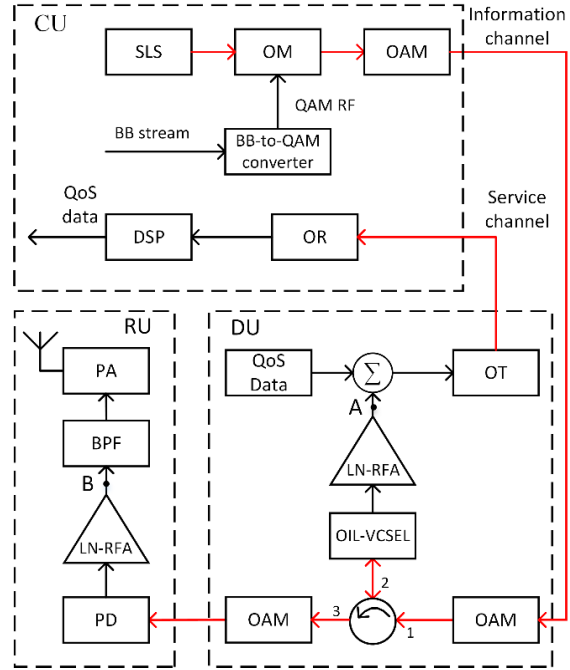

Figure 2. Block-diagram of the proposed duplex communication channel for NG-RAN, where SLS, OM, OAM, DSP, OR, OT, PD, LN-RFA, BPF, and PA stand for semiconductor laser source, optical modulator, optical amplifier, digital signal processor, optical receiver, optical transmitter, photodetector, low-noise RF amplifier, bandpass RF filter, and power RF amplifier (Optical connections are painted in red, electrical connections – in black)

## 4. PROOF-OF-CONCEPT EXPERIMENT

To validate the proposed concept a laboratory experiment was realized where a wafer-fused long-wavelength vertical cavity surface emitting laser of RTI Research, LLC in the form of a chip, optically injection locked by a master laser, was used. All measurements were carried out on the Probe Station EP6 from Cascade Microtech using coplanar RF probe and fiber-optics probe. A detailed description of the research object is given in [8]. The purpose of the experiment is to confirm the functionality and effectiveness of the block-diagram of a duplex communication channel for NG-RAN described in Section 3, when optically transmitting a high-speed digital signal with multi-position QAM of a RF carrier. Note that during the experiment the OIL-VCSEL operates in forward DC biased mode without switching to reverse DC bias in photodetector mode as required with a standard pin-photodiode, which is ensured by optical injection locking [8].

Figure 3 shows the testbed for proof-of-concept experiment, the layout of which is based on Figure 2 with the exclusion of non-essential for the confirmation elements after points A and B. The testbed contains pin-photodiode (Finisar, BPDV2150: 43-GHz bandwidth, 0.6-A/W responsivity), a pair of low-noise RF amplifiers (Mini-circuits ZX60-542LN-S+: 4.4-5.4-GHz frequency band, 24-dB gain, 1.9-dB noise figure), and two coils of single-mode fibers SMF-28+, as well as measuring tools including Vector Signal Generator (Keysight MXG N5182B) and 4-channel Mixed Signal Oscilloscope (Keysight Infinium MSOS804A).

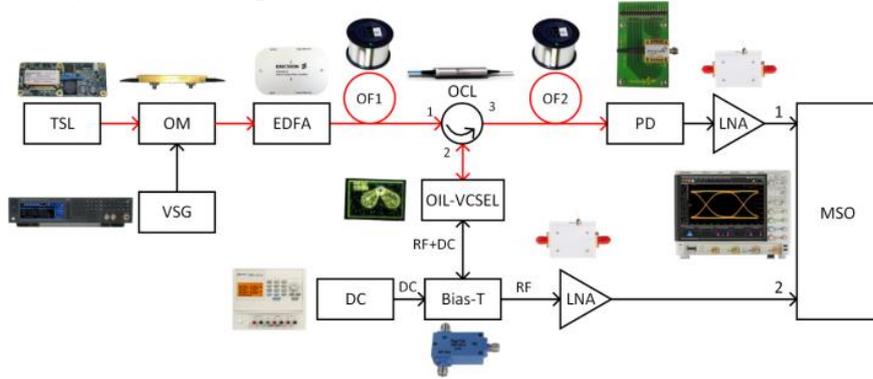

Figure 3. Testbed for the proof-of-concept experiment where TSL, OM, EDFA, OCL, OF, PD, LNA, DC, VSG, and MSO stand for tunable semiconductor laser (master laser), optical modulator, erbium-doped fiber amplifier, optical circulator, optical fiber, photodetector, low-noise amplifier, DC source, vector signal generator, and mixed signal oscilloscope, respectively. (Optical connections are painted in red, electrical connections – in black)

Figure 4 presents the results of the experiment, where optical carrier of the frequency near 192.2 THz is intensity modulated by the 560 Mbps 64-QAM RF signal of 5 GHz. Namely, in Figures 4 (a, b), MSO RF spectra are shown at the inputs 1 and 2, correspondingly. Figure 4 (c) shows the constellation diagram at the input 1 or 2 of the MSO, and Figure 4 (d) shows EVM values vs OF1 length at the inputs 1 and 2 of the MSO.

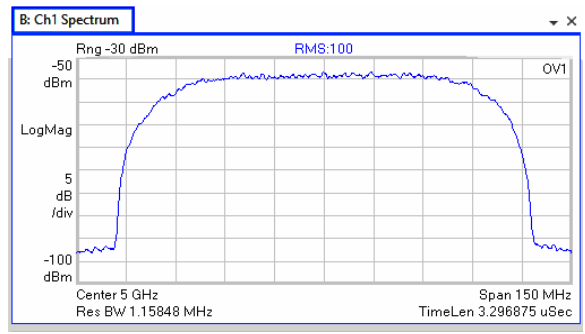

(a) RF spectrum at the input 1 of the MSO

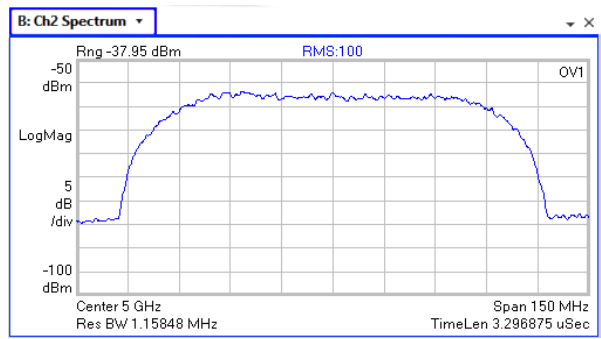

(b) RF spectrum at the input 2 of the MSO

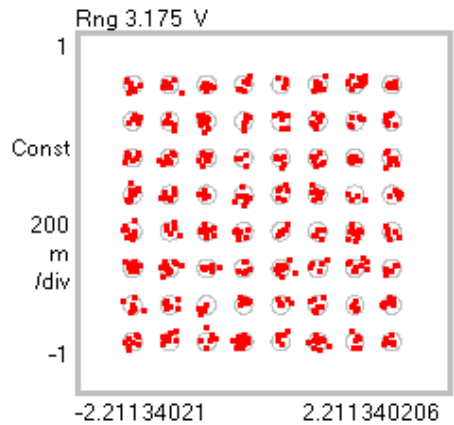

(c) The constellation diagram at the input 1 or 2 of the MSO

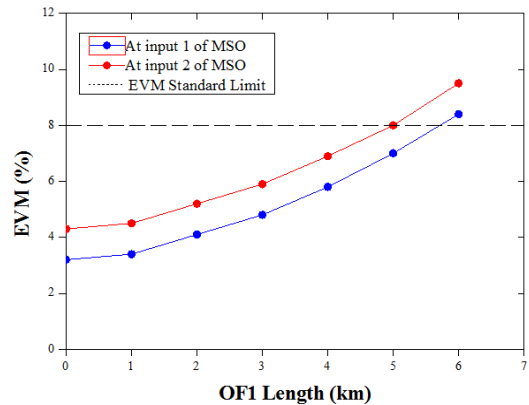

(d) EVMs vs OF1 length at the inputs 1 and 2 of the MSO

Figure 4. The results of the proof-of-concept experiment

The outcome that can be drawn from the above results is the following. In the DU under test, the values of error vector magnitude in back-to-back configuration are 3.2% for downlink channel to RU and 4.3% for uplink channel to CU, which is significantly less compared to the standard EVM threshold for 64-QAM of 8% [20]. This maximum acceptable transmission quality is achieved, when the distance between CU and DU is increased up to 5 km, which is quite realistic in a practical 5G access network based on small cell concept.

## 5. CONCLUSIONS

In this paper, a new cost-efficient concept to realize real-time monitoring of quality-of-service metrics and other service data in 5G and beyond access networks using a small-cell architecture is proposed and discussed. The essence of the proposed solution is investigated in detail on the example of a block diagram of a duplex communication channel, including a microcell base station and a functionally split picocell base station containing a central unit, a distribution unit, and a set of remote radio units each connected via optical fiber link. A distinctive feature of the proposed approach, which provides an improvement in the power and cost characteristics, is to use in the distribution unit, a vertical cavity surface emitting laser in the optical injection locked mode that simultaneously operating as an optical transmitter and as a resonant cavity enhanced photodetector. Thanks to this, it becomes possible to transmit information data to distribution unit via the downlink channel and real-time service data related to the status, quality of service, etc. via the uplink channel for processing them at the central unit. The feasibility and efficiency of the proposed solution are confirmed by a proof-of-concept experiment when optically transmitting a high-speed digital signal with 64-position quadrature amplitude modulation of a 5-GHz radio-frequency carrier, which is widely exploited in access networks of fifth-generation cellular communication systems based on Radio-over-Fiber technology and small cell architecture scenario.

The further research in this direction will focus on a detailed studying the VCSEL-based optical transmission with real-time monitoring function for ongoing 5G and beyond access networks of millimeter-wave band. The fundamental feasibility of this path due to optical injection locking has already been considered in a number of scientific publications, for example, in [21] and confirmed experimentally [22] using the same VCSEL chip as in this study.

**Authors**

Prof. Dr. Mikhail E. Belkin - received an engineering degree in radio and television from Moscow Institute of Telecommunications, in 1971, Ph. D. degree in telecommunication and electronic engineering from Moscow Technical University of Telecommunications and Informatics, in 1996, and Sc. D. degree in photonics and optical communications from Moscow State Technical University of Radio-Engineering, Electronics and Automation, in 2007. The theme of his Sc. D. degree thesis is 'Analog Fiber Optic Systems with multiplexing on RF and Microwave Subcarriers and Multiservice HFC Networks on their Base'. He has written more than 250 scientific works in English and Russian. The major current R&D fields are fiber-optic devices and systems, microwave photonics, photonic ICs, incoming cellular communication networks, computer-aided design.

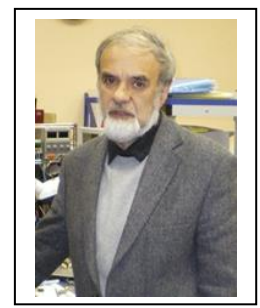

At present, M. E. Belkin is the Director of the Scientific and Technological Center "Integrated Microwave

Photonics", Professor of the department "Optical and Optoelectronic Devices and Systems", Institute of Physics and Technology, MIREA - Russian Technological University, and he is a member of IEEE's MTTS, LEOS (now PhS), and COMSOC from 2006, and a member of OSA from 2018.

Leonid Zhukov graduated from Moscow technical university of communications and informatics in 2019, and works at Russian Technological University MIREA 2016 as a laboratory assistant researcher after an engineer at the STC 'Integrated Microwave Photonics'. He is a co-author of 1 monographs published by IntechOpen. His main research interests are microwave-photonics, and telecommunication technologies.

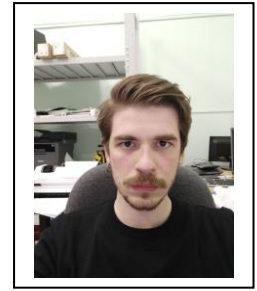

Professor Dr. Alexander. S. Sigov is an expert in Solid State Physics and Electronics. He contributed extensively to the phenomenology of magnets, ferroelectrics and multiferroics, physics of ferroic-based heterostructures, thin films, etc. The results of his scientific activity are reflected in more than 300 papers, reviews, book chapters, 19 monographs and textbooks, including the well-known "Defects and Structural Phase Transitions" together with A. Levanyuk. For many years, he has chaired the Department of Nanoelectronics in MIREA. He created his own school, inspiring and mentoring many talented scientists. In 2006, he was elected a Member of the Russian Academy of Sciences. He is the head of the Russian Academy Council on Dielectrics and Ferroelectrics, member of numerous scientific societies, Associate Editor of international journals Ferroelectrics and Integrated Ferroelectrics, Editor and member of Boards of more than ten Russian national journals, Chair of the Council on Physics and Astronomy of the Russian Foundation for Basic Research. At present, Alexander Sigov is the Head of Nanoelectronics Dept. and President of MIREA – Russian Technological University, Moscow, Russia, Doctor of Physics, and Fellow member of the Russian Academy of Sciences.

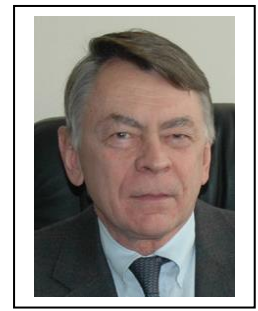